\documentclass[useAMS,usenatbib,usegraphicx,letters]{mn2e}
\usepackage{epsf}

\title[Lens galaxy environments and anomalous flux ratios]
{Lens galaxy environments and anomalous flux ratios in gravitational lenses}
\author[M. Oguri]
{Masamune Oguri\thanks{E-mail:oguri@astro.princeton.edu} \\
Princeton University Observatory, Peyton Hall,
Princeton, NJ 08544, USA.\\
Department of Physics, University of Tokyo, Hongo
7-3-1, Bunkyo-ku, Tokyo 113-0033, Japan.}
\begin{document}

\date{\today}

\voffset- .65in

\pagerange{\pageref{firstpage}--\pageref{lastpage}} \pubyear{2004}

\maketitle

\label{firstpage}

\begin{abstract}
The fraction of substructures required to account for anomalous flux
ratios in gravitational lens systems appears to be higher than
that predicted in the standard cold dark matter cosmology. We present a
possible alternative route to anomalous flux ratios from lens galaxy
environments. We consider the compound lens system such that a lens
galaxy lie in a group or cluster, and estimate the contribution of
substructures in the group/cluster to the fraction using an analytic
model of substructures. We find that the contribution becomes dominant
when the impact parameter of the lens is less than $\sim 30\%$ of the
virial radius of the group/cluster. This indicates that the
environmental effect can partly explain the high incidence of anomalous
flux ratios.  
\end{abstract}


\section{Introduction}

It appears quite common that observed flux ratios between gravitationally 
lensed quasar images are difficult to match while image positions can be 
fitted easily with simple smooth mass models. \citet{mao98} argued these
anomalous flux ratios as evidence for substructures in lens galaxies. 
If this is true, it serves as a powerful probe of substructures in
galactic halos which may not be luminous enough
\citep{metcalf01,chiba02,dalal02,keeton03,bradac04}. For instance, 
\citet{dalal02} concluded that substructures should  comprise
$0.6\%-7\%$ (90\% confidence) of the mass of typical lens galaxies in
order to explain flux ratios in several lens systems. Since the Cold
Dark Matter (CDM) model predicts $5\%-15\%$ of mass in substructures
\citep[e.g.,][]{delucia04,weller05}, 
they argued that the result strongly supports for the CDM model.
Not all anomalous flux ratios can be ascribed to the simpleness of
the smooth part of lens potentials \citep{evans03,kawano04} or any
propagation effects because the degree of the (de-)magnification in
observations depends on image parities \citep[e.g.,][]{kochanek04}. 

However, a caveat is that the fraction of substructures is in practice
a very sensitive function of distances from the halo center: Both
numerical simulations \citep[e.g.,][]{delucia04} and analytic models
\citep[e.g.,][]{oguri04} clearly demonstrate that substructures
preferentially lie in outer parts of dark halos. This means that 
very little substructure is projected within the Einstein radius, since 
the Einstein radius is usually much smaller than the virial radius of
halos. Thus the fraction of substructures in the CDM model appears to be  
{\it smaller} than that required to explain anomalous flux ratios
in gravitational lenses if we take the spatial distribution of
substructures into account \citep{evans03,mao04}. \citet[][ see also
\citealt{chen03}]{metcalf05} claimed that the discrepancy may be
resolved by considering extragalactic halos along line of sight.
Another possible explanation is some contributions from stellar
microlensing. For instance, \citet{chiba05} took mid-infrared images 
of B1422+231 and PG1115+080, and concluded that the anomalous flux ratio
of PG1115+080 is likely to be caused by stellar microlensing.

In this {\it Letter}, we present a possible alternative route to
anomalous flux ratios. Specifically, we consider the compound lens system
such that a lens galaxy lies in a group or cluster, and discuss to what
extent substructures in the group/cluster could contribute to cause
anomalous flux ratios. Although the situation is not universal, recent
studies indicate that such compound systems (including incidence of 
foreground groups) appear to be quite common
\citep{keeton00,blandford01,keeton04,fassnacht04,oguri05}. 
Observations imply that many of them lie in groups, and lens systems in
clusters are less common. The effect of correlated matter outside the
lens halo was studied by \citet{chen03}, however they considered typical
field galaxies as lens objects, thus the situation is somewhat
different from ours. Throughout the Letter we assume a Lambda-dominated
cosmology with the mass density parameter $\Omega_{\rm m}=0.3$, the
vacuum energy density parameter $\Omega_{\Lambda}=0.7$, the
dimensionless Hubble constant of $h=0.7$, and the rms fluctuation
normalization $\sigma_8=0.9$.  The results presented here will be,
however, depend little on the specific choice of cosmological parameters.

\section{Mass Fraction of Substructures from An Analytic Model}

Following \citet{mao04}, we compute the fraction of the projected
surface density, $f_{\rm sub}$, to discuss the efficiency of flux
anomalies. To do so, we use an analytic model of substructures developed
by \citet{oguri04}. The model takes account of two dominant dynamical
processes that drive dominantly subhalo evolution: one is mass-loss
caused by tidal interaction with the host halo, and the other is the
orbital decay caused by dynamical friction which drives massive subhalos
to the inner part of the host halo. The model also considers the formation
epoch variation of the host halo, and the orbital decay of satellite
halos outside the host halo virial radius. It was shown that the derived
analytic distributions agree well with the results of recent
high-resolution N-body simulation \citep{delucia04}. Although the model
may be less realistic than semi-analytic approaches
\citep{bullock00,hayashi03,taylor04,zentner05,penarrubia05}, it has the
advantage of being able to compute spatial and mass distributions very
easily. 

The fraction of the projected surface density $f_{\rm sub}$ is defined
by the projected mass in substructures within an annulus with radii
$R_-$ and $R_+$ divided by the projected total mass within the
annulus. We choose $R_-$ and $R_+$  so that the annuli are equally
spaced in $\log(R/r_{\rm vir})$ with a step size 0.2, where $r_{\rm
vir}$ is a virial radius of a halo that corresponds to a lens galaxy.
First, we compute  $f_{\rm sub}$ in an isolated halo with mass $M$. The
mass inside a cylinder with radius $R$ are computed as 
\begin{equation}
M(<R)=\frac{4\pi}{3}\int_0^1 d(\cos\theta) \rho(<r_{\rm max};M)r_{\rm
 max}^3,
\label{eq:mass}
\end{equation}
\begin{equation}
 r_{\rm max}=\min\left(r_{\rm vir},\frac{R}{\sin\theta}\right),
\end{equation}
where $\rho(<r;M)$ is the average density within a sphere of radius $r$,
and $\theta$ is an angle with respect to the projection direction.
We can calculate the density profile of substructures $\rho_{\rm sub}$
and corresponding mass $M_{\rm sub}(<R)$ by integrating the
distributions of substructures \citep{oguri04} over the mass of
substructures. Then the mass fraction of substructures is
\begin{equation}
f_{\rm sub}=\frac{\Delta M_{\rm sub}(R_-,R_+)}{\Delta M(R_-,R_+)}.
\label{eq:fsub1}
\end{equation}
where the mass within annuli is denoted as $\Delta M(R_-,R_+)=M(<R_+)-M(<R_-)$.

\begin{figure}
\begin{center}
 \includegraphics[width=0.9\hsize]{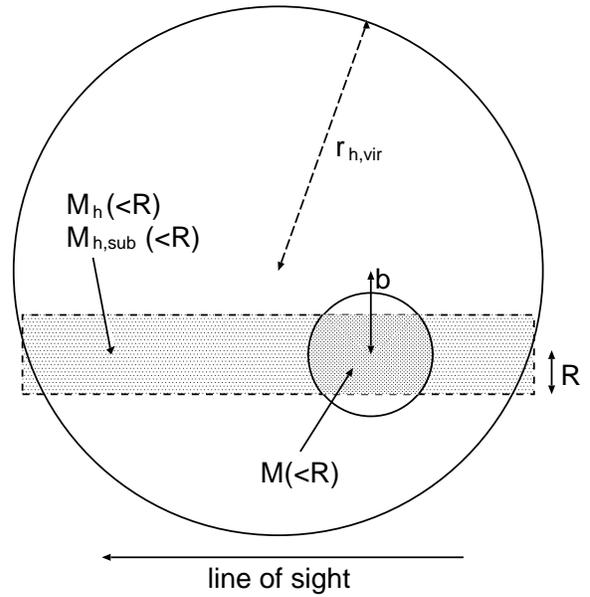}
\end{center}
\caption{Calculation of $f_{\rm sub}$ for a compound lens. The projected
 masses $M(<R)$, $M_{\rm h}(<R)$, and $M_{\rm h,sub}(<R)$ are given by 
equations (\ref{eq:mass}), (\ref{eq:mh}), and (\ref{eq:mhs}), respectively.
\label{fig:conf}}
\end{figure}

Next we consider the situation that a lens galaxy lies in a group or
cluster. In this case, the lens galaxy should correspond to a substructure
(hereafter referred as primary substructure) in a host halo with mass
$M_{\rm h}$. Then to compute the lens potential we have to consider the
mass associated the host halo (group or cluster) as well as the mass of
the substructure. Here we consider the situation that another
substructure in the host halo causes anomalous flux ratios in
gravitational lenses (hereafter referred as secondary substructure).
To quantify this contribution, we compute $f_{\rm sub}$ in this
compound lens system. When $R$ is much smaller than the virial
radius of the host halo, the mass of dark matter and substructures
associated with the host halo can be approximated as  
\begin{eqnarray}
 M_{\rm h}(<R) & \approx &  2\pi R^2\int_b^{r_{\rm h,vir}}\rho(r;M_{\rm h})\frac{r}{\sqrt{r^2-b^2}}dr,\label{eq:mh}\\
 M_{\rm h,sub}(<R) & \approx &  2\pi R^2\int_b^{r_{\rm h,vir}}\rho_{\rm sub}(r;M_{\rm h})\frac{r}{\sqrt{r^2-b^2}}dr,\label{eq:mhs}
\end{eqnarray}
with $r_{\rm h,vir}$ being the virial radius of the host halo with mass
$M_{\rm h}$ and $b$ being the impact parameter of the primary
substructure (see Figure \ref{fig:conf}). 
Then the mass fraction from substructures in the host halo is
\begin{equation}
 f_{\rm sub}=\frac{\Delta M_{\rm h,sub}(R_-,R_+)}{\Delta
  M(R_-,R_+)+\Delta M_{\rm h}(R_-,R_+)},
\label{eq:fsub2}
\end{equation}
where $\Delta M(R_-,R_+)$ is the mass associated with the primary
substructure calculated from equation (\ref{eq:mass}). Although the
primary substructure suffers from tidal stripping which makes the size
of the substructure smaller, we neglect this because the contribution of 
the outer part to the mass estimate is quite small.

\section{Result}

\begin{figure}
\begin{center}
 \includegraphics[width=0.8\hsize]{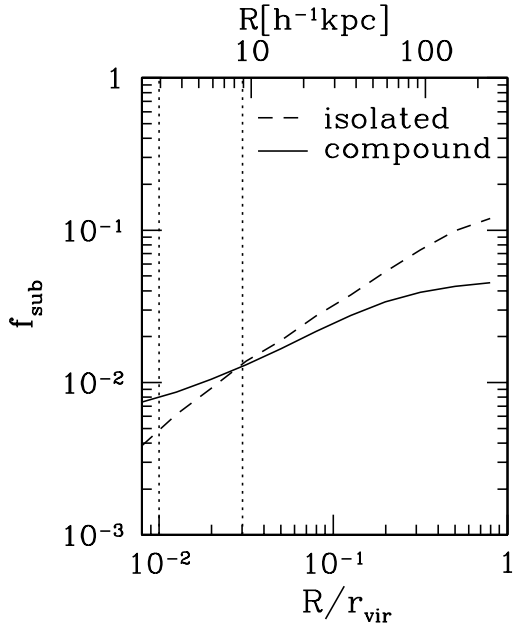}
\end{center}
\caption{The mass fraction of substructures $f_{\rm sub}$ as a function
 of projected radius $R$. The fractions of an isolated lens
 (eq. [\ref{eq:fsub1}]) and compound lens (eq. [\ref{eq:fsub2}])  are
 plotted by solid and dashed lines, respectively. 
 We assume the lens (sub-)halo with $M=5\times
 10^{12}h^{-1}M_\odot$, the host halo with $M_{\rm
 h}=10^{14}h^{-1}M_\odot$, the impact parameter of the lens subhalo in the
 host halo $b/r_{\rm h,vir}=0.2$. They are placed at $z=0.3$.
 The vertical dotted lines indicate the position of typical lensed
 images, $R/r_{\rm vir}=1\%-3\%$ \citep{mao04}.  
\label{fig:fsub}}
\end{figure}

We calculate $f_{\rm sub}$ in the following setup: The mass of the
isolated halo or the primary substructure, which corresponds to the lens
object, is $M=5\times 10^{12}h^{-1}M_\odot$. As a redshift, we choose a
typical lens redshift of $z=0.3$. For the compound system, we set the
mass of the host halo $M_{\rm h}=10^{14}h^{-1}M_\odot$ (i.e., 
$M_{\rm h}/M=20$). The virial radii of the lens and host halos are 
$r_{\rm vir}=295h^{-1}{\rm kpc}$ and $r_{\rm h,vir}=800h^{-1}{\rm kpc}$,
respectively. The primary substructure is placed so that the impact
parameter becomes $b/r_{\rm h,vir}=0.2$. The value of the impact
parameter is slightly smaller than the average impact parameter
calculated from the radial distribution of substructures assuming 
the ratio of substructure to host halo masses to be $10^{-2}$ (see
Figure 3 of \citealt{oguri04}), $\bar{b}/r_{\rm h,vir}\sim 0.3$,  but is
a reasonable value if we take account of the bias due to the convergence
of the host halo (see \S \ref{sec:sum}). We integrate the mass of 
substructures in the range $10^{-4}<m/M<10^{-1}$ ($5\times
10^8h^{-1}M_\odot<m<5\times 10^{11}h^{-1}M_\odot$) to compare our results
with those of \citet{mao04}. However, almost universal form of the mass
function of substructures \citep{oguri04} assures that relative amount of 
$f_{\rm sub}$ between isolated and compound lens systems does
not change very much even if we shift the range to lower masses. 

\begin{figure}
\begin{center}
 \includegraphics[width=1.0\hsize]{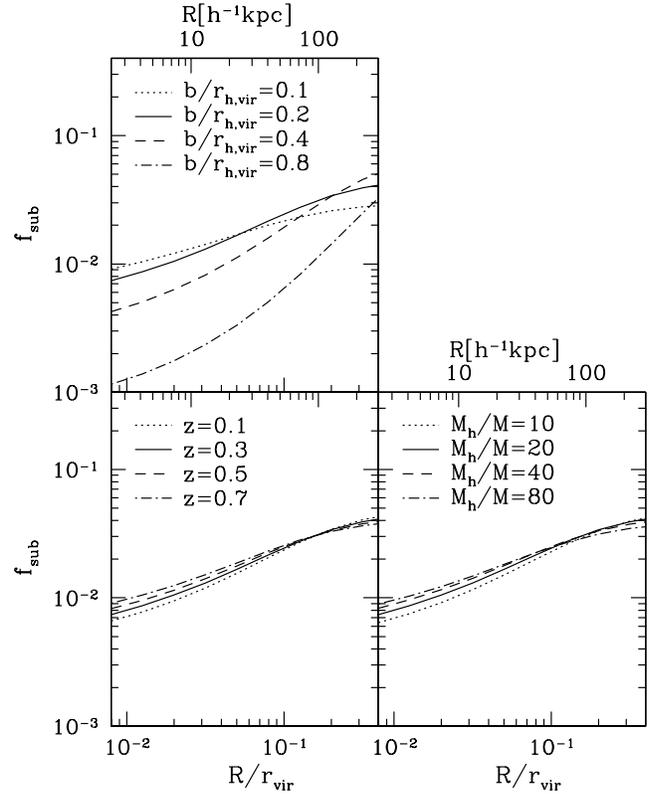}
\end{center}
\caption{Dependences of $f_{\rm sub}$ for the compound lens on
 several parameters: The impact parameter $b$ ({\it upper left}), the 
 redshift $z$ ({\it lower left}), and the mass of the host halo $M_{\rm
 h}$ ({\it lower right}). We change each parameter by fixing the other
 parameters to the fiducial values. 
\label{fig:fsub_par}}
\end{figure}

Figure \ref{fig:fsub} shows $f_{\rm sub}$ as a function of the projected
radius $R$. We plot $f_{\rm sub}$ for both isolated compound systems.
We note that for the compound system we show $f_{\rm sub}$ from host
group/cluster only: The lens galaxy in group/cluster itself should
also have substructures that will increase $f_{\rm sub}$ of the compound
lens system further. 
For the isolated halo, we reproduce the result of \citet{mao04}. We find
that $f_{\rm sub}$ for the compound lens is comparable to, or even
higher than, that for the isolated lens at the position of typical
lensed images, $R/r_{\rm vir}=1\%-3\%$. To examine how the result is
dependent on our specific setup describe above, in Figure
\ref{fig:fsub_par} we show the dependence of $f_{\rm sub}$ for the
compound lens on several parameters. We find that $f_{\rm sub}$ is
quite sensitive to the impact parameter $b$ but rather insensitive to
other parameters including the mass of the host halo.  Therefore we
conclude that substructures in the host halo of the compound lens could
be a dominant source of anomalous flux ratios when the impact parameter
is $b/r_{\rm h,vir}\la 0.3$. The predicted values, however, are still
$f_{\rm sub}\la 10^{-2}$ (without baryons) and thus  they do not
completely eliminate, though do reduce, the discrepancy between
predicted and observed fraction of substructures. 

The expected level of the contribution from substructures in a
group/cluster can also be estimated by a simple argument. In our
fiducial parameter set, the convergence at image position ($R/r_{\rm
vir}=1\%-3\%$) is $\kappa\sim 0.25$ (assuming a source redshift
$z_{\rm s}=1$), and the external convergence is $\kappa_{\rm env}\sim
0.05$. On the other hand, at $20\%$ of a virial radius the fraction of
substructures is $\sim 0.06$ (see Figure \ref{fig:fsub}). From these
numbers, $f_{\rm sub}$ is estimated as 
\begin{equation}
 f_{\rm sub}(R/r_{\rm vir}=1\%-3\%)\sim\frac{0.06\times \kappa_{\rm env}}{\kappa+\kappa_{\rm env}}\sim 0.01.
\end{equation}
The estimated number is roughly consistent with the result presented in
Figure \ref{fig:fsub}. The external convergence $\kappa_{\rm env}$ is a
strong function of the impact parameter, thus the reason that $f_{\rm
sub}$ is sensitive to the impact parameter $b$ can be also understood
from this expression.

\section{Summary and Discussion}\label{sec:sum}

We have examined the impact of lens galaxy environments on substructure
lensing. Using an analytic model of substructures constructed by
\citet{oguri04}, we have computed the fraction of substructures in the
situation that a lens galaxy lies in a group or cluster. Specifically,
we have regarded a subhalo in a halo as a lens galaxy, and
estimated the contribution of other substructures in the halo 
to the fraction of substructures which determines the efficiency of 
substructure lensing. We have found that the fraction depends mainly on
the impact parameter of the primary substructure (lens galaxy); the
contribution becomes dominant when the impact parameter is less than
$\sim 30\%$ of the virial radius of the host halo. We emphasize that 
the effect is rather insensitive to the mass of the host halo; our
result suggests that even a poor group could have a great impact on
substructure lensing. Therefore we conclude that lens galaxy
environments are important when we discuss anomalous flux ratios. 
This, in turn, means that lens systems in dense environments may be 
more desirable sites for detecting dark halo substructures, with either
anomalous flux ratios or more direct methods
\citep[e.g.,][]{yonehara03,inoue05}. 

The result appears to contradict with that of \citet{chen03} who showed 
that nearby correlated structure is highly sub-dominant compared to
substructure within the lens halo. However, they assumed average
galaxies for lens objects and estimated the number density of clumps
outside the lens using two-point correlation function $\xi(r)$, $n\sim
\bar{n}(1+\xi(r))$, where $\bar{n}$ is the average number density of 
clumps in the universe. In contrast, we have assumed rather specific
environment such that lens objects correspond to galaxies in a group or
cluster, thus the situation is somewhat different from theirs. 
In our case, it is more appropriate to estimate the number density 
with the number density of clumps inside the group/cluster,
$\bar{n}_{\rm G}\sim \Delta_{\rm vir}\bar{n}$ (where $\Delta_{\rm
vir}\sim 300$ is the characteristic value of the overdensity of
collapsed objects), instead of $\bar{n}$. This means that the number
density of clumps in our situation should be higher by a factor $\sim
\Delta_{\rm vir}$, and this extra factor accounts for the difference
between our and their results.  

The importance of this new route may be enhanced further if we consider
the effect of baryon. As \citet{mao04} noted, the baryon infall makes
the density profile of lens galaxies steeper, $\rho\propto r^{-2}$,
and hence substructures will be disrupted more easily. However, the
argument is not applied to compound lens systems because the impact of
baryon cooling on groups and clusters is much less significant.

\begin{table}
 \caption{Summary of known environments of lens systems with anomalous
 flux ratios. The sample is constructed from \citet{keeton03} and
 \citet{kochanek04}. Note that the anomalous flux ratios of some of
 these system may be caused by stellar microlensing rather than lensing
 by substructures.\label{table:sub}} 
 \begin{tabular}{@{}ll}
  \hline
  Name &  Environment\\
  \hline
  MG0414+0534 & - \\
  B0712+472 & group \citep{fassnacht02}\\
  RX J0911+0551 & cluster \citep{bade97}\\
  SDSS J0924+0219 & -\\
  PG 1115+080 & group \citep{kundic97}\\
  RX J1131-1231 & -\\
  B1422+231 & group \citep{tonry98}\\
  B1933+503 & - \\
  B2045+265 & group \citep{fassnacht04}\\
  \hline
 \end{tabular}
\end{table}

It is important to address how ubiquitous such lens systems
are. \citet{keeton00} predict that $\sim~25\%$ of lens galaxies are in
groups or clusters, and \citet{blandford01} argued that the fraction
could be even higher. Although \citet{premadi04} and \citet{dalal05}
found somewhat smaller environmental effect, it appears that the models  
predict significantly smaller values for external shear than those
obtained by modeling strong lensing data. In addition, it should be
taken into account that anomalous flux ratios have been discussed
primarily using quadruple lens systems. This indicates the sample would
be biased significantly toward the higher fraction of the compound
systems, because environmental effects enhance the quadruple to double
ratio \citep{keeton04}. Another effect we have to think of is the bias
associated with the convergence of the group or cluster which increases
lensing cross sections by a factor $\sim (1-\kappa)^{-2}$. These effects
preferentially select compound systems with small impact parameters,
indicating that the compound lens system considered in this paper may be
quite ubiquitous.  Indeed, the survey of environments of lens systems
with strong flux anomalies, which is summarized in Table
\ref{table:sub}, shows that more than half of the lens systems lie in
groups/clusters, implying a correlation between flux anomalies and dense
galaxy environments. 

\section*{Acknowledgments}
The author thanks Jounghun Lee, Neal Dalal, and Premana Premadi for 
useful discussions and comments, and the anonymous referees for
suggestions. The author is supported by JSPS through JSPS Research
Fellowship for Young Scientists.  


\label{lastpage}

\end{document}